\documentclass[twocolumn,showpacs,preprintnumbers,amsmath,amssymb,prb]{revtex4-1}
\usepackage{amsmath}
\usepackage{amssymb}
\usepackage{graphicx}
\usepackage{dcolumn} 
\usepackage{bm} 
\usepackage{longtable}
\usepackage[usenames]{color}
\usepackage[normalem]{ulem}
\usepackage[usenames]{color}
\usepackage[normalem]{ulem}

\begin{document}
\title{Interband spin-orbit coupling between anti-parallel spin states \\ in Pb quantum well states}
\author{Bartosz Slomski$^{1,2}$, Gabriel Landolt$^{1,2}$, Stefan Muff$^{1,2}$, Fabian Meier$^{1,2}$, J\"urg Osterwalder$^1$ and J. Hugo Dil$^{1,2}$}
\address{$^{1}$Physik-Institut, Universit\"at Z\"urich, Winterthurerstrasse 190,
CH-8057 Z\"urich, Switzerland}
\address{$^{2}$Swiss Light Source, Paul Scherrer Institut, CH-5232 Villigen,
Switzerland}
\date{\today}
\begin{abstract}
Using spin and angle-resolved photoemission spectroscopy we investigate a momentum region in Pb quantum well states on Si(111) where hybridization between Rashba-split bands alters the band structure significantly. Starting from the Rashba regime where the dispersion of the quasi-free two-dimensional electron gas is well described by two spin-polarized parabolas, we find a breakdown of the Rashba behavior which manifests itself (i) in a spin splitting that is no longer proportional to the in-plane momentum and (ii) in a reversal of the sign of the momentum splitting. Our experimental findings are well explained by including interband spin-orbit coupling that mixes Rashba-split states with \emph{anti-parallel} rather than parallel spins. Similar results for Pb/Cu(111) reveal that the proposed hybridization scenario is independent on the supporting substrate.
\end{abstract}

\pacs{73.21.Fg, 71.70.Ej, 79.60.Dp}

\maketitle

\section{Introduction}
The study of the spin structure of non-magnetic low dimensional systems by means of spin- and angle-resolved photoemission (SARPES) has intesified in recent years \cite{Dil:2009R}. One of the driving forces has been the discovery of systems with a large Rashba-type spin splitting on a variety of three-dimensional \cite{Ishizaka:2011,Landolt:2012}, two-dimensional \cite{Ast:2007,Yaji:2010,Hopfner:2012} and even one-dimensional systems \cite{Wells:2009,Okuda:2010, Tegenkamp:2012}. The fact that for topological insulators the topology of the bulk bands is reflected in the spin structure of the surface states \cite{Hasan:2010} provided a further boost to this type of experiments. Also here a variety of SARPES experiments have been performed in order to elucidate the details of the spin structure and the variations from a purely in-plane helical spin system \cite{Hsieh:2009,Hsieh:2009N,Xu:2011,Souma:2011,Eremeev:2012}. In most of these cases the initial state has been considered as a pure spin state although under the influence of strong spin-orbit interaction the spin is no longer a good quantum number. Recently it has been shown that due to the orbital mixing in the topological surface state of Bi$_{2}$Se$_{3}$ also the spin structure of this state consists of mixed spin states \cite{Cao:2012arXiv}. 

Also for Rashba systems deviations from the pure spin scenario are expected to occur when spin-orbit coupling (SOC) induced avoided crossing hybridization leads to an opening of energy gaps between Rashba-split bands. 
When bands belonging to the same group representation approach one another, SOC opens a hybridization gap, resulting in an avoided crossing of bands that is absent within a nonrelativistic treatment. Close to the gap the Bloch eigenstates get mixed and form new eigenstates. Since the driving mechanism for the mixing of the states is SOC whose operator ($\xi \textbf{L} \cdot \textbf{S}$) contains both a spin-conserving and a spin-flip term  \cite{Ebert:1997}, we would like to understand how the SOC-induced hybridization alters the properties of an initially Rashba-split 2DEG.

To date, only few experiments have investigated the subtle interplay between the orbital momentum and spin of hybridized states using the direct method of spin- and angle-resolved photoemission spectroscopy. For example A.G. Rybkin \emph{et al.} \cite{Rybkin:2012} found that quantum well states (QWS) in the low-Z material aluminum deposited on the high-Z substrate W(110) acquire
large spin splittings through the interaction with spin-polarized interface states because of the avoided crossing hybridization. In Ag films grown on Si(111) and covered with the $\sqrt{3}$-Bi-Ag alloy, it has been shown that spin-selective hybridization occurs between spin-degenerate Ag quantum well states and strongly Rashba-type spin split alloy surface states \cite{He:2010,Frantzeskakis:2008}. In both studies the authors observed a hybridization between states with parallel spin direction. In contrast to this, first-principles calculations and ARPES measurements of Bi/Cu(111) \cite{Mirhosseini:2009} and BiAg$_2$ \cite{Bentmann:2012} indicate a hybridization between states with \emph{anti-parallel} spin direction. Whether the former (hybridization between parallel spins) or the latter (hybridization of \emph{anti-parallel} spins) scenario is present in a system depends on the orbital symmetry of the involved wave functions.

As a model system we investigate QWS in an ultra-thin Pb film grown on the Bi reconstructed Si(111) substrate. The studied energy-momentum region contains a single and pronounced electron-like band spin-split by the Rashba effect which starts to deviate from the parabolic dispersion at larger wave vectors. The electronic and structural properties of Pb QWS on Si(111) have been intensively studied in the past. The reason for the large interest is because Pb on Si(111) shows a rich variety of phenomena ranging from anomalies in the superconducting transition temperature \cite{Guo:2004, Brun:2009,Eom:2006}, the formation of magic height islands \cite{Budde:2000}, extremely long excited-state lifetimes \cite{Kirchmann:2010}, structure dependent Schottky barriers \cite{Heslinga:1990, Ricci:2004}, interface dependent effective mass \cite{Slomski:2011}, and Rashba-type spin splitting \cite{Dil:2008, Slomski:2011B}. Here we add the new phenomenon of interband spin-orbit coupling between Rashba-type spin-split QWS of opposite spin directions. 
 
By means of SARPES combined with a vectorial spin analysis \cite{Meier:2009NJP} we identify an interband spin-orbit coupling caused by the spin-flip term of the SOC operator. Our findings are well matched by a simple analytical model adapted from Ref. \cite{Bentmann:2012}. Similar behavior is observed experimentally in Pb QWS grown on Cu(111).

\section{Experiment}
The (S)ARPES experiments were performed with the COPHEE setup at the Surface and Interface Beamline at the Swiss Light Source of the Paul-Scherrer-Institut \cite{Hoesch:2002}. The energy and angular resolution of the spectrometer in the spin-integrated (spin-resolved) ARPES measurements were set to 30 (80) meV and 0.5$^{\circ}$ (1.5$^{\circ}$), respectively. All valence band photoemission data were recorded at a photon energy of 24 eV due to the optimal photoemission cross-section \cite{Dil:2004}, with p--polarized light.
The base pressure of the chamber was below 2$\times$10$^{-10}$ mbar and the sample temperature during the measurements was T $<$ 80 K. The $n$-type Si(111) ($44-62$~$\Omega$cm) sample was degassed at 600 K for 24~h and flashed several times above 1300 K to remove adsorbates and to form the (7$\times$7) surface reconstruction \cite{Takayanagi:1980}. The cleanliness of the sample was checked with ARPES, where the surface states could be identified, and by inspecting the LEED pattern which showed the typical (7$\times$7) surface reconstruction with intense superstructure spots. The Bi interface with  $(\sqrt{3}\times \sqrt{3})$R$30^{\circ}$ symmetry and 1 monolayer (ML) coverage in substrate units (1 ML = 7.83 $\times$ 10$^{14}$ atoms/cm$^{2}$) [henceforth Bi-$\sqrt{3}$] was prepared through the deposition of approximately 3 ML of Bi from a water-cooled Knudsen cell onto the clean Si(111)-(7$\times$7) surface at low temperature and subsequent annealing until the $(\sqrt{3}\times \sqrt{3})$R30$^{\circ}$ surface reconstruction appeared in LEED \cite{Wan:91PRB}.
On this reconstructed substrate a thin crystalline Pb film was grown by depositing at a rate of $\frac{1}{3}$ ML per minute after cooling down again to below 100 K. Under these conditions Pb grows in a layer-by-layer mode allowing for a precise determination of the layer thickness through monitoring the binding energies of the subbands as a function of deposition time. 
The ultra-thin Pb film on Cu(111) was fabricated by the deposition of Pb from a water-cooled e-beam evaporator, at a pressure below $3 \times 10^{-10}$ mbar, onto a Cu(111) sample held at 80 K. The single crystal Cu(111) was cleaned before deposition by several cycles of Ar ion bombardment and annealing.

\section{The Rashba-Bychkov effect}
The Rashba model was originally proposed to describe the motion of an electron in a 2DEG subjected to the electric field due to a broken structural inversion symmetry perpendicular to the plane of confinement, i.e. $\mathbf{E} = E\mathbf{e_z}$. In such an environment the electric field acts as a magnetic field in the rest frame of a moving electron with momentum $\hbar \mathbf{k_{\parallel}}$ as represented by the Rashba Hamiltonian,
\begin{eqnarray}
\hat{H}_{RB} = \alpha_{RB}(\mathbf{e_z} \times \mathbf{k_{\parallel}})\cdot \mathbf{S}
\end{eqnarray}
Here $\alpha_{RB}\propto E_z$ is the so-called Rashba parameter which describes the strength of the interaction between the spin $\mathbf{S}$ of an electron and the effective magnetic field $\mathbf{B_{eff}} \propto (\mathbf{e_z}~\times~\mathbf{k_{\parallel}})$. 
The Rashba effect is therefore in close analogy to the Zeeman effect, except that the quantization axis for the spin, defined by $\mathbf{B_{eff}}$, is a function of both, the momentum of the electron and the electric field.
Owing to the Rashba effect the initially two-fold degenerate parabolic dispersion splits into two parabolas with oppositely spin-polarized states:
\begin{eqnarray}
E^{\pm}(\mathbf{k_{\parallel}}) = E_{\overline{\Gamma}} + \frac{\hbar^2}{2m^{\star}}\mathbf{k}^2_{\parallel} \pm \alpha_{\text{RB}}|\mathbf{k_{\parallel}}| 
\label{Eq:Rashba}
\end{eqnarray}
where $E_{\overline{\Gamma}}$ is the binding energy at $k_{\parallel} = 0$ \AA$^{-1}$ and $m^{\star}$ represents the effective mass of the electrons. 

The Rashba effect has been experimentally confirmed in many systems, such as in semiconductor heterostructures \cite{Nitta:1997}, in Shockley-type surface states on noble metals, e.g. Au(111) \cite{LaShell:1996,Nicolay:2001,Hoesch:2004} and recently in Cu(111) \cite{Tamai:2013}, in surface states of heavy metal surface alloys \cite{Ast:2007PRB, Ast:2007,Meier:2008,Ast:2008,Meier:2009PRB}, on semiconducting surfaces covered with heavy metals \cite{Gierz:2009, Yaji:2010}, and in one monolayer of Bi on Cu(111) \cite{Mathias:2010}. Many insights into the origin and the strength of spin-orbit interaction have been achieved by studying these systems \cite{HeinzmannDil:2012}. Nowadays it is accepted that the size of the Rashba effect in real systems is not a simple function of the electric field at the surface or interface, but that it is more sensitive to the details of the wave function close to the nuclei which in turn are related to structural properties of the surface.

\begin{figure}
\begin{center}
\includegraphics[width=0.5\textwidth]{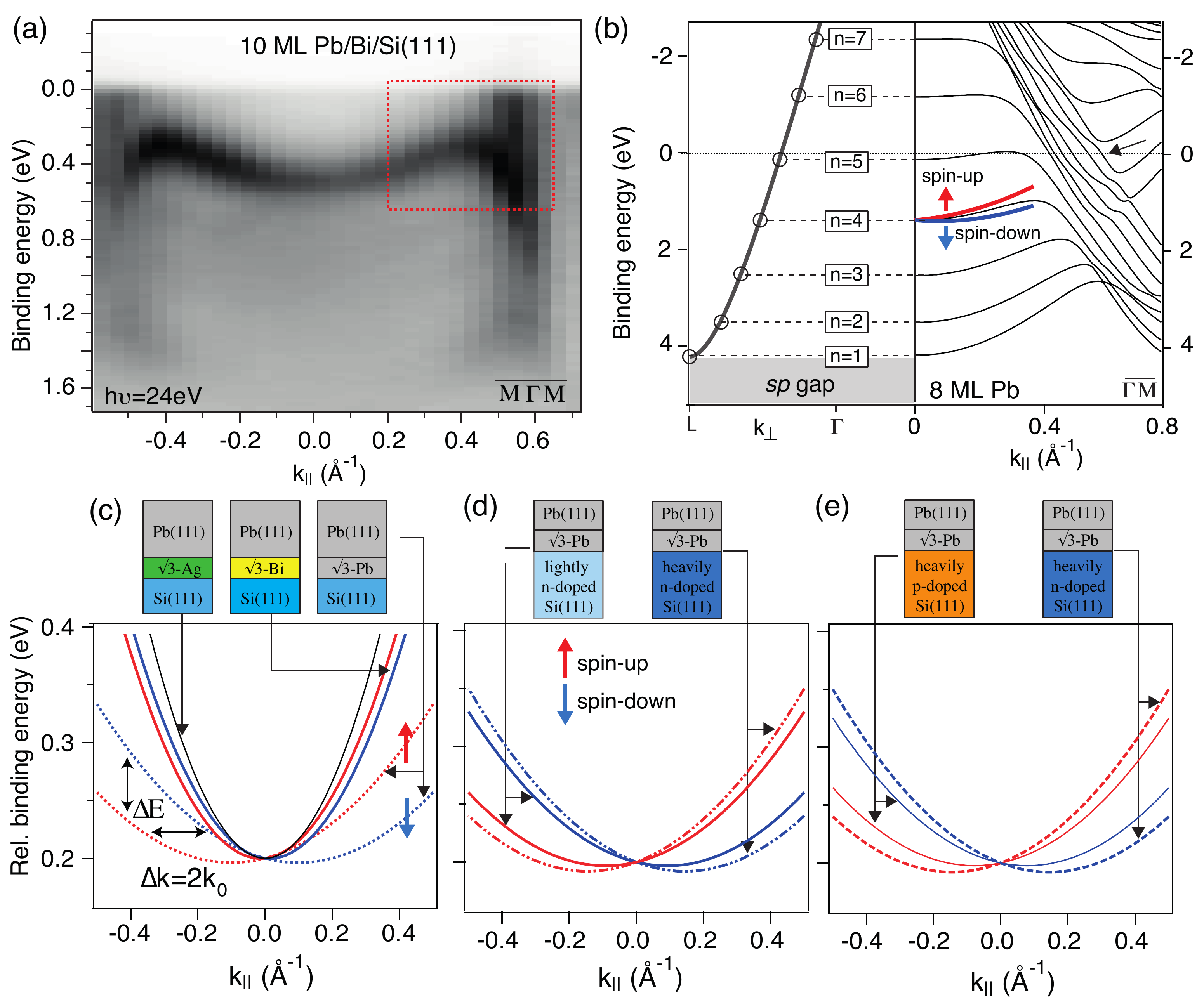}
\caption{(color online) (a) Electronic structure of 10 ML Pb on Bi-$\sqrt{3}$/Si(111) measured with angle-resolved photoemission spectroscopy. (b) Quantization of the band structure along $k_{\perp}$ as a consequence of size reduction along the growth direction (left panel). Calculated band structure of a 8 ML free-standing Pb film (right panel). The subband with the quantum number $n=4$ is schematically spin-split into two spin-polarized parabolas with opposite spin direction according to the Rashba effect. (c-e) Calculated spin-split parabolas using parameters ($\alpha_{RB}$ and $m^{\star}$) as deduced from the (S)ARPES experiment to illustrate the sensitivity of the Rashba effect on (c) the interface chemistry, (d) the donor concentration, and (e) the type of dopant. Note that the band dispersions are displayed on a relative binding energy scale to illustrate also the differences in the effective mass as a function of interface.}
\label{Fig1}
\end{center}
\end{figure}

A nice playground for this model is the Rashba effect in Pb QWS \cite{Dil:2008, Slomski:2011B}.
Figure \ref{Fig1}(a) shows the in-plane electronic structure of an ultra-thin film formed by 10 monolayers of Pb grown on the Bi reconstructed Si substrate and measured with ARPES. The M-shaped QWS with a binding energy ($E_b$) $E_{b,\overline{\Gamma}}$ =  490 meV at normal emission ($k_{\parallel}~=~0$\AA$^{-1}$) arises from the quantization of the band dispersion along the $\Gamma$-$\text{L}$ direction due to the size reduction which is illustrated in the left panel of Fig. \ref{Fig1}(b) by the example of a 8 ML Pb film. The corresponding in-plane band structure of the ultra-thin Pb film is shown in the right panel of Fig. \ref{Fig1}(b) and was calculated with density functional theory using the \emph{Wien2k} simulation package \cite{Schwarz:DFT} incorporating SOC. 

Our previous research focussed on the spin texture of the bands in the momentum region around the surface Brillouin zone (SBZ) center. Here the system is a good approximation of a quasi-free 2DEG confined in a quantum well with a structural inversion asymmetry. We could show that these states show a Rashba effect and that the size of the spin splitting is sensitive to the metal-substrate interface. In the first approach we have varied the chemistry of the interface by studying various interfactants such as Pb, Bi and Ag, which form highly regular structures of $\sqrt{3}$ symmetry on Si(111). It is found that the size of the Rashba effect changes dramatically between these interfaces: replacing the Pb interface by a Bi layer reduced the Rashba parameter by 60\%, whereas QWS in Pb films grown on a Ag reconstructed Si substrate showed no measurable spin splitting \cite{Slomski:2011B}, see Fig. \ref{Fig1}(c). The sensitivity of the Rashba parameter to the donor concentration of the Si substrate revealed a very promising pathway toward the realization of a spintronic device: by increasing the donor concentration by a factor of 20, we could tune the Rashba parameter by a factor of two \cite{Slomski:2013}, see Fig. \ref{Fig1}(d). A strong sensitivity of the Rashba parameter is also found when changing the type of dopant, while keeping the dopant concentration almost constant (Fig. \ref{Fig1}(e)). These changes of the Rashba effect are well understood by a modified charge density distribution within the Pb film as a consequence of an interface and doping dependent Schottky barrier \cite{Slomski:2013}.

\section{Interband Spin-Orbit Coupling in Pb QWS}
\begin{figure*}[htb]
\begin{center}
\includegraphics[width=0.85\textwidth]{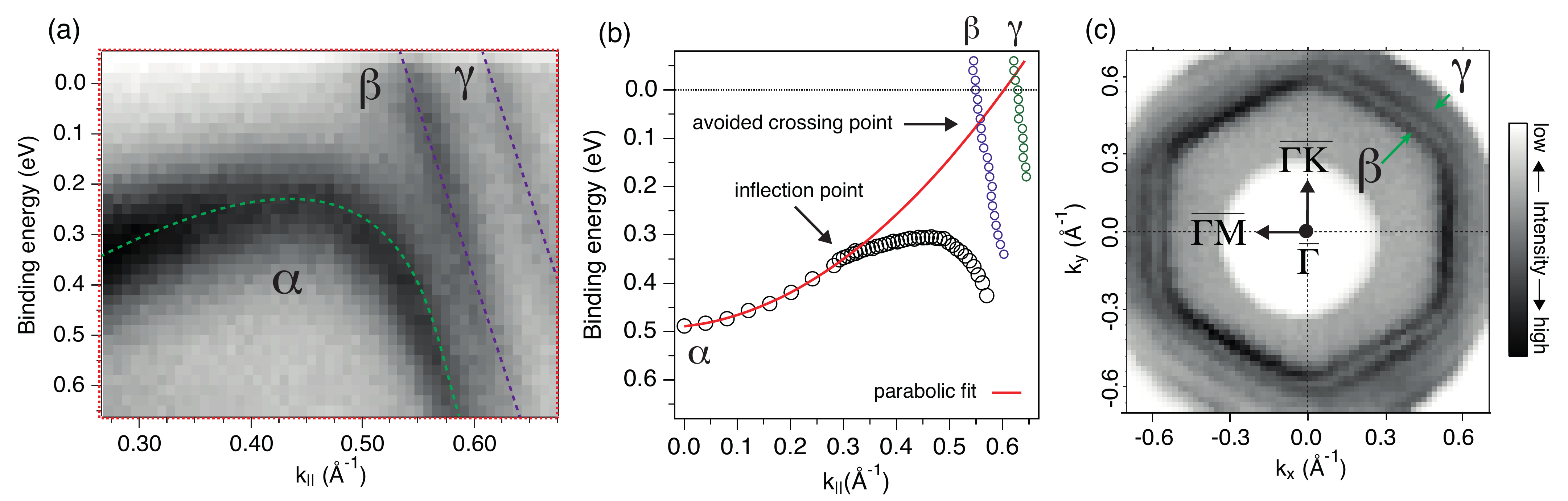}
\caption{(color online) (a) Band structure region where the SOC-induced avoided crossing hybridization leads to the pronounced deviation from the parabolic dispersion of the $\alpha$-band. (b) Extracted dispersion of the $\alpha$, $\beta$ and $\gamma$ band. The parabolic fit gives an effective mass of $m^{\star}=3.2$ $m_e$ and starts to deviate from the actual dispersion at $\approx$~0.3 \AA$^{-1}$. (c) Fermi surface of the ultra-thin Pb film (10 ML) with the high-symmetry direction $\overline{\Gamma}$-$\overline{M} \parallel k_x$ and $\overline{\Gamma}$-$\overline{K} \parallel k_y$. }
\label{Fig1b}
\end{center}
\end{figure*}

The main focus of the present study is the outer region in the band structure marked by the dashed box in Fig. \ref{Fig1}(a) and enlarged in Fig. \ref{Fig1b}(a). It consists of three bands labeled as $\alpha$, $\beta$, $\gamma$, which are derived from Pb $6p$ orbitals  because the binding energies lie above the Pb $sp$ symmetry band gap in the $\Gamma$-$\text{L}$ direction (cf. Fig. \ref{Fig1}(b)). More specifically, states that are described by an upward parabolic dispersion with band minimum at $\overline{\Gamma}$ are mainly of $6p_z$ symmetry (magnetic quantum number $m_l = 0$), whereas states with downward dispersion and band maximum at $\overline{\Gamma}$ are of $6p_{x,y}$ symmetry $(m_l \pm 1)$. 
Within a momentum range of $k_{\parallel} = \overline{\Gamma} \pm 0.3$ \AA$^{-1}$, the dispersion of the $\alpha$-band is well described as a Rashba system via Eq. \ref{Eq:Rashba} with an effective mass of $m^{\star} = 3.2$ $m_e$ \cite{Slomski:2011} and a Rashba constant of $\alpha_{RB}$~=~0.033 eV\AA\ \cite{Slomski:2011B}. 
Remarkably, the initially $6p_z$ derived $\alpha$-band changes its curvature from being electron-like ($m^{\star} > 0$) to hole-like ($m^{\star} < 0$) for in-plane momenta $|k_{\parallel}| > 0.3$ \AA$^{-1}$, which is a first indication of hybridization \cite{Upton:2005}. Such a pronounced change in the electronic dispersion has been also observed for monolayers of Pb on graphitized SiC and explained as due to SOC-induced hybridization and is therefore related to the peculiar band structure of the Pb film \cite{Dil:2007}. In Fig. \ref{Fig1b}(b) we plot the  dispersion of the $\alpha$-band, which was obtained by fitting energy distribution curves utilizing Voigt profiles, as a function of momentum. It is seen that the actual electron-like dispersion starts to deviate from the experimental data at $k_{\parallel} \approx 0.32$ \AA$^{-1}$ [hereafter termed the inflection point]. For clarity we have also included the dispersions of the hole-like $\beta$ and $\gamma$ bands, respectively. The parabolic fit of the $\alpha$-band would intersect the $\beta$-band at $(E_b, k_{\parallel})$ $\approx$ (100 meV, 0.55 \AA$^{-1}$) [hereafter termed the avoided crossing point] in case the SOC-induced hybridization were turned off. The Fermi level crossing of the $\beta$- and $\gamma$-band is also nicely resolved in Fig. \ref{Fig1b}(c) which displays the Fermi surface (FS) of the ultra-thin Pb film. The hexagonal shape of the FS reflects the $(111)$ crystal symmetry with corners pointing into the high symmetry direction $\overline{\Gamma}$-$\overline{\text{K}}$. Furthermore, by following e.g. the inner $6p_{x,y}$ band on the FS we observe an intensity modulation with a periodicity of $\frac{2}{3}\pi$, in line with the 3-fold rotational symmetry $(C_{3\nu}$) of the crystal structure.

\begin{figure*}[t]
\begin{center}
\includegraphics[width=0.6\textwidth]{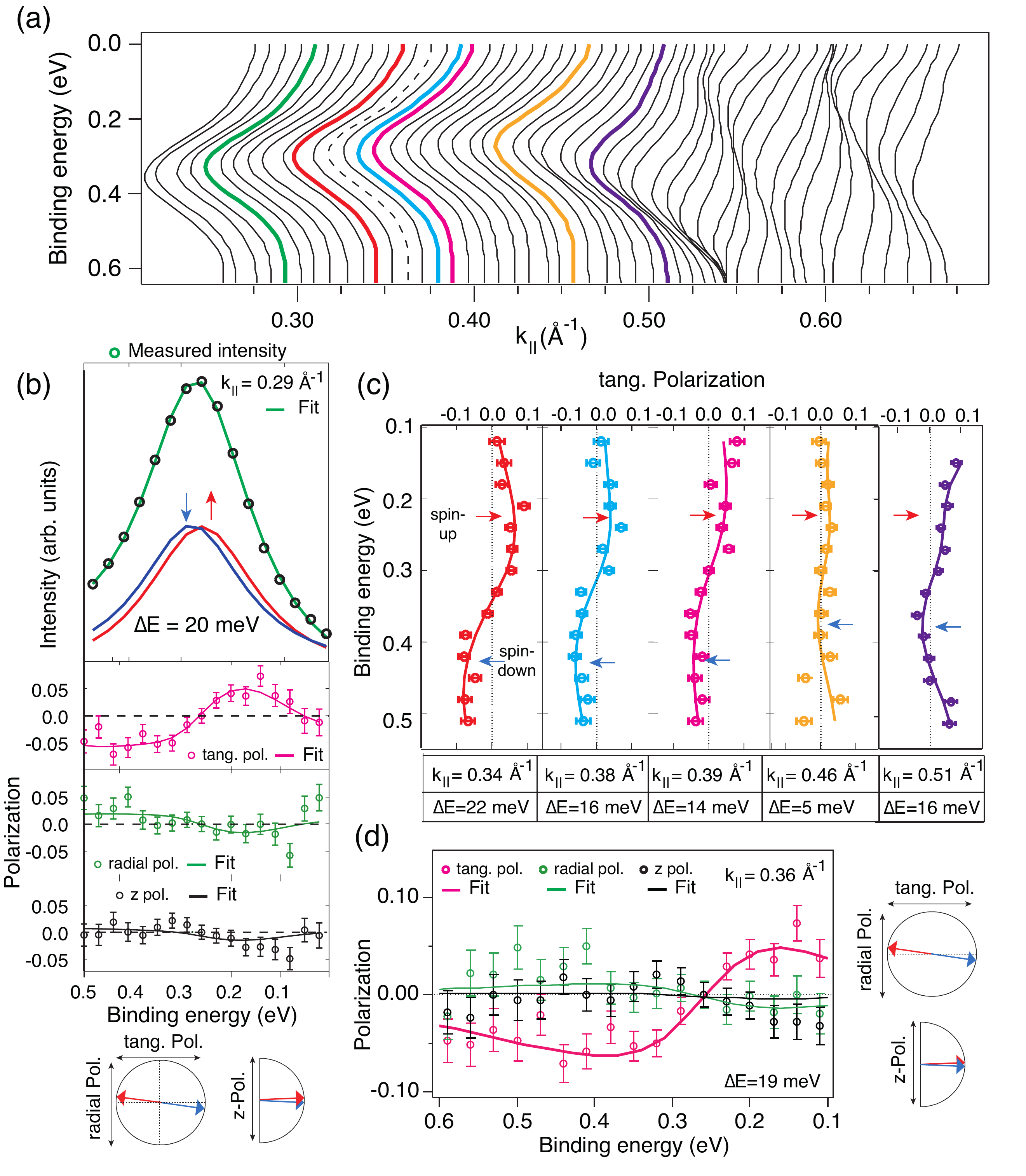}
\caption{(color online) (a) Energy distribution curves of the $\alpha$-band as a function of $k_{\parallel}$. (b) (upper panel) Measured spin-integrated EDC at $k_{\parallel}$ = 0.29 \AA$^{-1}$ with Voigt intensity profiles obtained from the two-step fitting routine. (middle panel) Corresponding spin polarization data and fits in the tangential, radial and z-direction. (lower panel) Spin polarization vectors revealed from the fit. (c) Tangential polarization data and fits at increasing in-plane momenta toward the avoided crossing point. (d) Measured spin polarization data and fits (left panel) and polarization vectors (right panel) in the hybridization regime ($k_{\parallel} = 0.36$ \AA$^{-1}$).} 
\label{Fig2}
\end{center}
\end{figure*}

Motivated by the peculiar dispersion of the $\alpha$-band we first study the evolution of its spin splitting as a function of momentum, starting in the Rashba regime.  
In Fig. \ref{Fig2}(a) we show spectra recorded as a function of $k_{\parallel}$ extracted from Fig. \ref{Fig1b}(a). We start with a spin-resolved energy distribution curve (SR-EDC) taken at $k_{\parallel} = 0.29$ \AA$^{-1}$ using the two-Mott detector scheme, which is sensitive to all three spin polarization components $(P_x, P_y, P_z)$ \cite{Hoesch:2002} which are defined as the expectation value of the spin operator, i.e $P_i = \langle\Psi |S_i|\Psi\rangle$ with $i=x,y,z$.

At this particular momentum, far away from the region where the $\alpha$-band bends downward, we expect a Rashba-like spin splitting and spin orientation. The data and the corresponding results using the self-consistent vectorial spin analysis \cite{Meier:2008} are shown in Fig. \ref{Fig2}(b). We observe up-down excursions in the tangential spin polarization $(P_{\tan})$, the direction perpendicular to the momentum, and a spin splitting of $\Delta E$ = 20 meV ($\Delta E := E_b^{\downarrow} - E_b^{\uparrow}$) between the spin-up and spin-down states. The polarization along the radial direction is below the detection limit of the spectrometer. An out-of-plane spin polarization ($P_z$), which has been reported for this system \cite{Slomski:2011B}, is suppressed because we measured along the $\overline{\Gamma}$-$\overline{\text{M}}$ direction, i.e. in the mirror plane of the $(1\times 1)$ SBZ. 

In the following we switch to the single Mott scheme and focus on the tangential spin polarization component. This way, we can decrease the $\Delta k$ steps for similar overall acquisition time and precisely follow the evolution of the spin splitting along the dispersion. Figure \ref{Fig2}(c) shows exemplary $P_{\tan}$ data and fits obtained from SR-EDCs taken at increasing in-plane momenta. The color coding of the data corresponds to that in Fig. \ref{Fig2}(a). We observe first an increased polarization amplitude and spin splitting toward $k_{\parallel}$ = 0.34 \AA$^{-1}$, followed by a reduction toward $k_{\parallel} = 0.5$ \AA$^{-1}$ (see also Fig. \ref{Fig8}(a)).

In the region where the $\alpha$-band bends downward the spin splitting seems to recover again with the same sign of the spin splitting, i.e. $\Delta E = 2\alpha_{RB}\cdot k_{\parallel} > 0$, which indicates a sign reversal in the momentum splitting $\Delta k = 2k_0$. As will be shown later, this sign change is better resolved in spin-resolved momentum distribution curves (MDC) rather than SR-EDCs. At this point we notice that in the region beyond the inflection point the tangential spin alignment persists and neither radial nor out-of-plane rotations of the spin polarization vector are observed. This is exemplified in Fig. \ref{Fig2}(d) which shows the 3D spin polarization data measured at $k_{\parallel}$~=0.36 \AA$^{-1}$. The spin polarization vectors of the two Rashba-split states are very similar to those displayed in Fig. \ref{Fig2}(b). 
 
\begin{figure*}[htb]
\begin{center}
\includegraphics[width=0.8\textwidth]{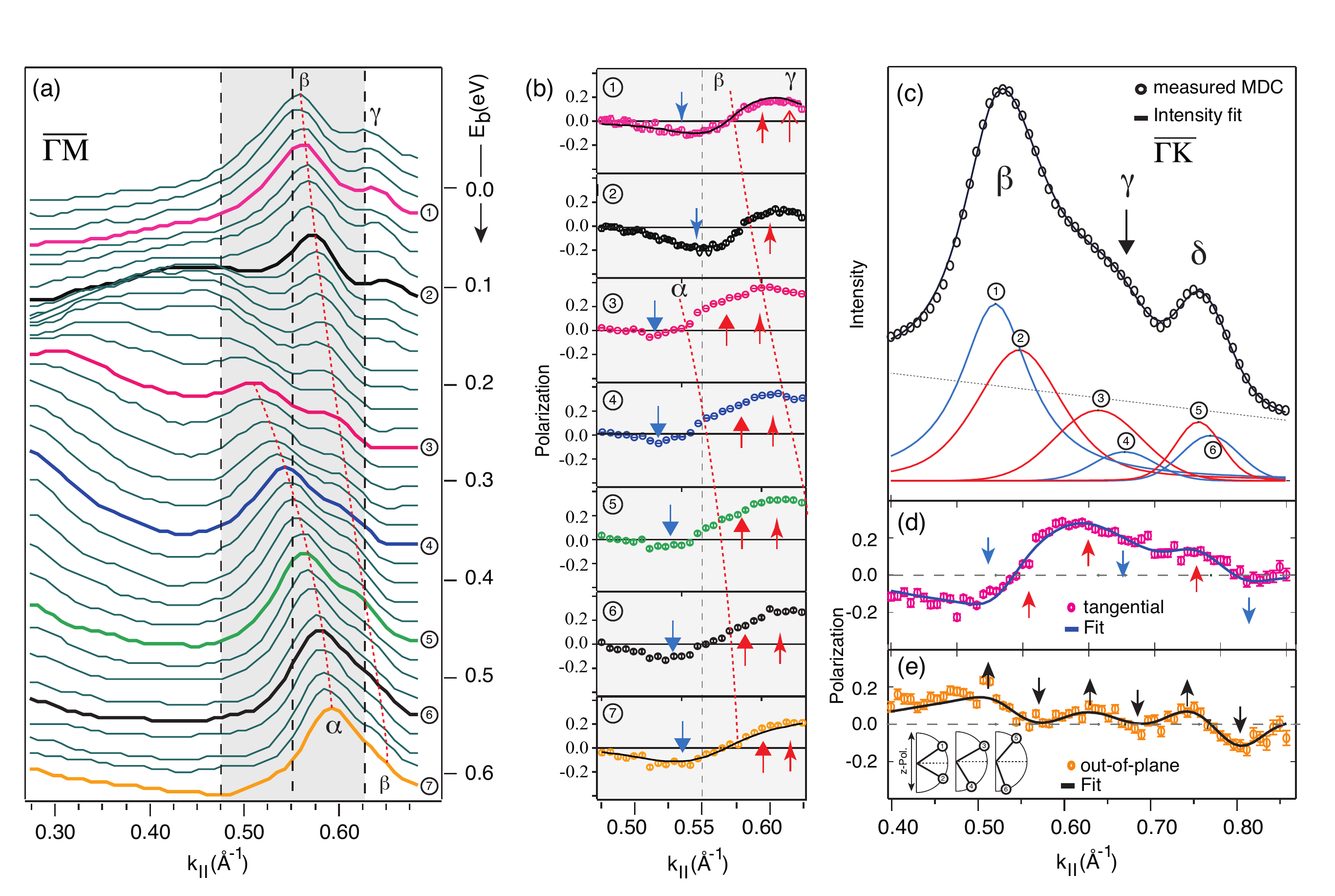}
\caption{(color online) (a) Spin-integrated momentum distribution curves along $\overline{\Gamma}$-$\overline{\text{M}}$ as a function of binding energy. (b) Measured tangential spin polarization data at various values of $E_b$; the numbers correspond to those in (a). (c) Measured total intensity MDC at the Fermi level along $\overline{\Gamma}$-$\overline{\text{K}}$ and corresponding spin polarization data and fits along the tangential direction (d)  and the z-direction (e). The inset of (e) shows the out-of-plane rotation of the spin polarization vectors at increasing in-plane momentum obtained from the self-consistent two-step fit.} 
\label{Fig3}
\end{center}
\end{figure*}

\begin{figure}[htb]
\begin{center}
\includegraphics[width=0.5\textwidth]{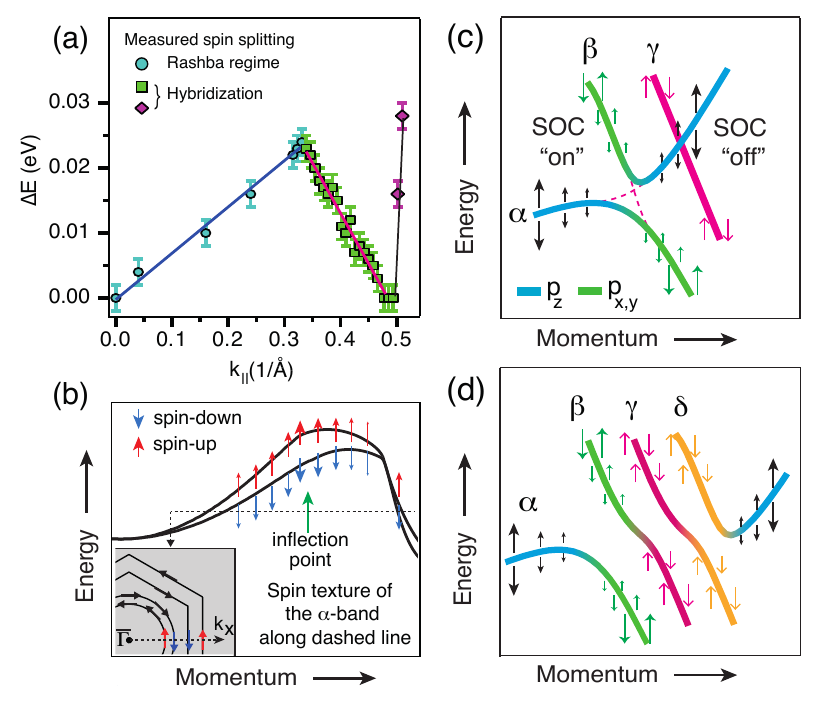}
\caption{(color online) (a) Measured k-dependent energy splittings of the $\alpha$-band over the full investigated momentum range, and linear fits. (b) Schematic drawing of the band splitting deduced from our fitting procedure to the experimental data. The inset shows the spin texture along the dashed line across the constant energy surface, illustrating the sign reversal of the momentum splitting. (c) Schematic drawing of the resulting dispersion of the $\alpha$- and the $\beta$-band when SOC-induced hybridization is taken into account. When the hybridization is turned off, the $\beta$-band continues with an electron-like dispersion and crosses the $\gamma$-band. (d) Resulting schematic dispersion of the bands after hybridization is turned on between the $\alpha$- and the $\beta$-band, between the $\beta$- and the $\gamma$-band, and between the $\gamma$- and the $\delta$-band.} 
\label{Fig8}
\end{center}
\end{figure}

Next, we focus on the spin texture of the $\alpha$- and the $\beta$-band by analyzing SR-MDCs. Figure \ref{Fig3}(a) shows spin-integrated MDCs extracted from Fig. \ref{Fig1b}(a). Starting from the high binding energy side, we observe maximum spectral weight on the downward dispersing $\alpha$-band and a small shoulder at the high momentum tail from the $\beta$-band. With decreasing $E_b$ the spectral weight shifts from the $\alpha$- to the $\beta$-band. Closer to $E_F$ both the $\beta$ and $\gamma$ bands can be clearly distinguished. Figure \ref{Fig3}(b) shows a series of spin polarization data from SR-MDCs. The polarization curve at $E_b \approx 0.6$ eV (scan~7) reveals a momentum splitting of the $\alpha$-band, that is in line with data taken in the EDC mode (cf. Fig. \ref{Fig2}(c) and Fig. \ref{Fig8}(b)), i.e. $k_{\alpha,\parallel}^{\downarrow} < k_{\alpha,\parallel}^{\uparrow}$ translates into $\Delta E > 0$. This substantiates our finding that the sign of the momentum splitting of the $\alpha$-band reverses upon passing the avoided crossing point. Because for this band both the sign of $m^{\star}$ and the sign of $\Delta k$ changes, the Rashba constant maintains its sign since $\alpha_{RB} = \hbar^2\Delta k / 2m^{\star}$.

However, for the adjacent $\beta$-band we find a sign reversal of the Rashba constant. This is seen from a comparison between the measurements at $E_b \approx$ 0.6 eV and $E_F$ (Fig.~\ref{Fig3}(b)). Scan 7 indicates that the spin-up peak (indicated by the second up-arrow at $k_{\parallel} \approx$ 0.64 \AA$^{-1}$) is located at lower $k_{\parallel}$ than the corresponding spin-down peak (not measured). In contrast, the analysis of scan 1 reveals that the spin-down peak of the $\beta$-band is at lower momentum ( $k_{\parallel} \approx$ 0.53 \AA$^{-1}$) than the corresponding spin-up peak  ($k_{\parallel} \approx$ 0.59 \AA$^{-1}$). This means, in between $E_b = 0.6$ eV and $E_F$ the Rashba constant of the band changed its sign as it disperses toward $E_F$. Indeed, when we compare scan 3 (at $\approx$ 0.25 eV) with scan 2 (at $\approx$ 0.1 eV) we can localize this sign change, which is close to the avoided crossing point (cf. Fig. \ref{Fig1b}(b)). The behavior is therefore different from the one of the $\alpha$-band, where both the sign of the effective mass and the sign of the momentum splitting are changed.

To unravel how the spin textures of these bands interplay, we have performed a SR-MDC (single Mott) at the Fermi level in an extended momentum range and along the $\overline{\Gamma}$-$\overline{\text{K}}$ high symmetry direction. The results are shown in Figs. \ref{Fig3}(c-e). 
First of all, the larger measured momentum range allows to investigate the spin texture of three $p_{x,y}$ bands ($\beta, \gamma, \delta$) at the Fermi level. Figure \ref{Fig3}(c) shows the spin-integrated MDC, and in (d) and (e) we show tangential and out-of-plane polarization data together with our analysis, respectively. There is a clear difference between the $P_{z}$ and $P_{\text{tan}}$ data. While the z-polarization reveals a spin configuration of alternating up-down spin pairs, the tangential polarization shows down-up-up-down-up-down excursions. This means that the Rashba constant changes sign between the $\beta$-band and the $\gamma$-band.
Furthermore, it is found that the out-of-plane rotation of the spin polarization vector increases with the in-plane momentum, which is understood by the fact that states of mainly $p_{x,y}$ orbital symmetry are more sensitive to the in-plane gradient of the potential. 
This finding is nicely in line with the observations for the high-Z metal surface alloys, e.g. Bi/Ag(111) \cite{Meier:2008}. 

We start the discussion on the origin of the measured spin structure with the right panel of Figure \ref{Fig1}(b) which displays a band structure calculation of a free-standing 8 ML Pb film. The avoided crossings of the bands are well reproduced by taking SOC into account as already demonstrated in Ref. \cite{Dil:2007}. At larger momenta it is seen that the $p_{x,y}$ band that crosses $E_F$ at $k_{\parallel}$ $\approx$ 0.6 \AA$^{-1}$ (the crossing is marked by an arrow), continues with an electron-like dispersion, adapted from an electron-like band (originally the $\alpha$-band), as long as no hybridization with a neighboring $p_{x,y}$ band occurs. As will be shown below this property has an important influence on the spin texture of the downward dispersing bands in between. 

Figures \ref{Fig8}(a, b) summarize our main findings concerning the spin texture of the $\alpha$-band. 
Fig. \ref{Fig8}(a) displays the measured spin splittings of the $\alpha$-band for the complete investigated momentum range and in (b) we draw a schematic band dispersion based on our findings. In going away from the $\overline{\Gamma}$ point, the spin splitting increases with the in-plane momentum, in accordance with the Rashba model. In between $k_{\parallel}$ = 0.34 and 0.5~\AA$^{-1}$, the measured energy splitting is found to decrease with the in-plane momentum. Such a behavior is in sharp contrast to the Rashba model and was also observed in 1 ML Au on W(110) \cite{Varykhalov:2008b}. We conjecture that hybridization starts to develop at the inflection point, i.e. as the interaction between the $\alpha$- and the adjacent $\beta$-band increases the energy splitting decreases. Finally, beyond the avoided crossing point the spin splitting recovers with a steeper slope as a consequence of the smaller effective mass, since $\Delta E \propto \alpha_{RB} \propto 1/m^{\star}$. Furthermore we find that the relative spin helicities of the spin-split components of the $\alpha$-band are reversed upon passing the momentum region where the band bends downward. Figure \ref{Fig8}(b) displays schematically the spin texture along the dashed line. 

Figures \ref{Fig8}(c, d) explain qualitatively the hybridization mechanism as revealed from the SARPES measurements. The SOC-induced hybridization between the $\alpha$- and the $\beta$-band leads to a gap opening and a sign reversal in the effective mass of both bands. Close to the gap the states hybridize by mixing the orbital symmetry and the spin character of the corresponding wave functions. That way the $\alpha$-band adopts the spin and orbital character of the $\beta$-band, which leads to a reversal of the sign of the momentum splitting and of the effective mass.
On the other hand, beyond the hybridization gap the $\beta$-band continues to disperse with mainly $p_z$ character (i.e. $m^{\star} > 0$) adopted from the $\alpha$-band  and approaches the adjacent $\gamma$-band, as schematically drawn in Fig. \ref{Fig8}(c) for the case of no interaction between these bands. When we turn the hybridization on (Fig.~\ref{Fig8}~(d)), the $\beta$-band adopts the $p_{x,y}$ orbital symmetry and the spin character of the $\gamma$-band. Consequently, it disperses again hole-like and  with a relative order of momentum splitting given by the $\gamma$-band which leads to a hybridization-induced reversal of the Rashba constant. The same arguments apply for the next bands with $p_{x,y}$ symmetry, as long as other downward dispersing bands are present. 

\begin{figure*}[htb]
\begin{center}
\includegraphics[width=0.8\textwidth]{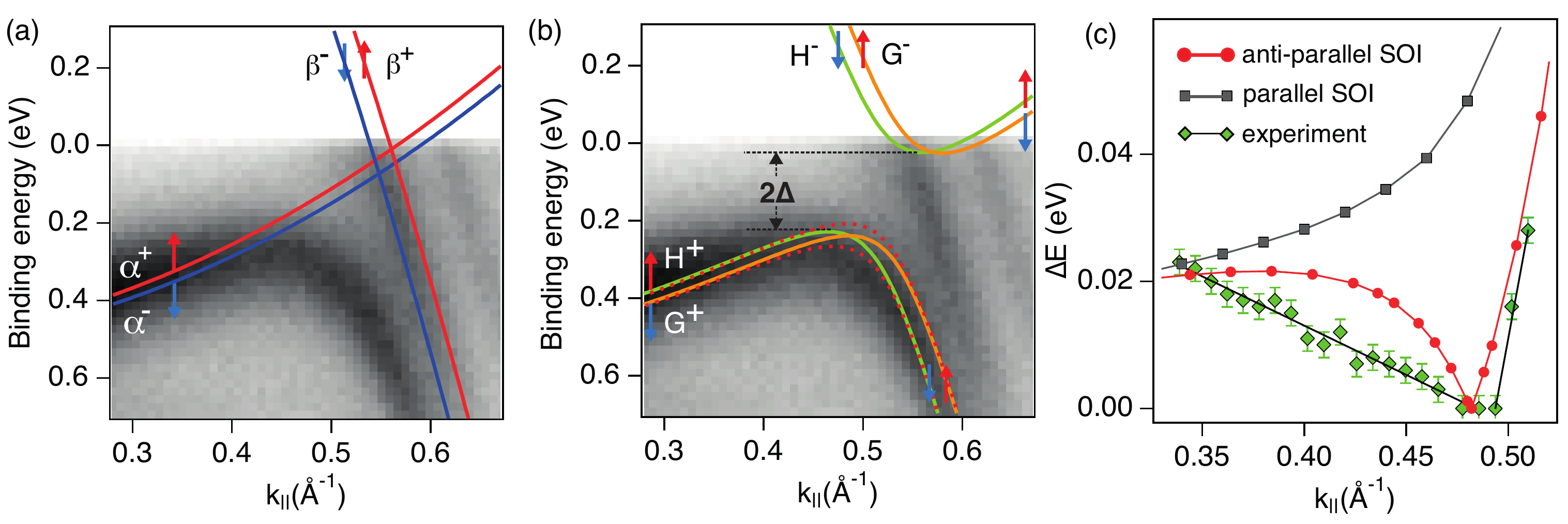}
\caption{(color online) (a) Comparison of the band structure in the avoided crossing region with the solution of the Rashba Hamiltonian. (b) The same region as in (a) overlaid by the new solutions of the model Hamiltonian incorporating SOC due to the Rashba effect and due the interband SO coupling for an interaction of strength $\Delta_{1,2}$ = 120 meV. (c) Energy splitting as a function of $k_{\parallel}$ in the hybridization region that starts at the inflection point. The measured spin splittings are compared with an interband SO coupling model that mixes states with parallel (black squares) and \emph{anti-parallel} (red circles) spins. A closer match to the experimental findings is achieved with \emph{anti-parallel-spin} interband SO coupling.} 
\label{Fig5}
\end{center}
\end{figure*}

To gain a quantitative understanding of the SOC-induced hybridization we adapt an interband spin-orbit coupling model suggested in Ref. \cite{Bentmann:2012} which was successfully used to explain the gap opening in the band structure of a strongly spin-split Rashba system formed in the surface alloy BiAg$_2$. This model assumes a hybridization between two Rashba branches with opposite spin direction. In the case of Pb/Bi/Si which represents a 2DEG with weak SOC due to the Rashba effect, we introduce the opening of two hybridization gaps with the matrix elements:
\begin{eqnarray}
\Delta_1 = \langle \alpha^{+}|H_{\text{SOC}}|\beta^{-}\rangle \hspace{1cm}
\Delta_2 = \langle \alpha^{-}|H_{\text{SOC}}|\beta^{+}\rangle 
\end{eqnarray}
Here $|\alpha^{\pm}\rangle$ ($|\beta^{\pm}\rangle$) denotes the eigenstates of the Rashba Hamiltonian. The modified energy dispersion is calculated via \cite{Bentmann:2012}
\begin{eqnarray}
H^{\pm} = \frac{1}{2}\left(\alpha^{+} +\beta^{-}\right) \pm \sqrt{\left(\frac{\alpha^+-\beta^-}{2}\right)^2 + \Delta_1^2} \\
G^{\pm} = \frac{1}{2}\left(\alpha^{-} +\beta^{+}\right) \pm \sqrt{\left(\frac{\alpha^--\beta^+}{2}\right)^2 + \Delta_2^2}
\end{eqnarray}
where $\alpha^{\pm}$ and $\beta^{\pm}$ denote the spin-up (+) and spin-down (-) energy branches. Close to the avoided crossing point the interband SO coupling mixes the states $|\alpha^{\pm}\rangle$ and $|\beta^{\pm}\rangle$ in new eigenstates:
\begin{eqnarray}
|H^{+}\rangle = h_k |\alpha^{+}\rangle + \sqrt{1-h^2_k}|\beta^{-}\rangle \\
|H^{-}\rangle = \sqrt{1-h^2_k}|\beta^{+}\rangle + h_k |\alpha^{-}\rangle  
\end{eqnarray}
and 
\begin{eqnarray}
|G^{+}\rangle = g_k |\alpha^{-}\rangle + \sqrt{1-g^2_k}|\beta^{+}\rangle \\
|G^{-}\rangle = \sqrt{1-g^2_k}|\beta^{-}\rangle + g_k |\alpha^{+}\rangle  
\end{eqnarray}
where $h_k$ and $g_k$ are k-dependent coefficients which determine the orbital and spin mixing ratio. Note that the new eigenstates are formed by a mixture of Rashba states with \emph{anti-parallel} spins. Before we discuss the modification of the $\alpha$-band dispersion induced by the interband spin-orbit coupling, we display in Fig. \ref{Fig5}(a) the eigenvalues of the $\alpha$ and $\beta$ bands according to the Rashba model. For the $\beta$-band we use a momentum splitting of $2k_0 = 0.021$ \AA$^{-1}$ as obtained from the vectorial spin analysis shown in Figs.~\ref{Fig3}(c-e) and an effective mass of $m^{\star} \approx - 0.5$ $m_e$. It is clear that this model alone can neither explain the sign reversal in the effective mass of the $\alpha$-band nor the modified spin structure of the $\alpha$- and the $\beta$-band across the avoided crossing point.

Figure \ref{Fig5}(b) shows the results of applying the interband SO coupling model, in addition to the Rashba-type spin-orbit interaction, on the energy dispersion of the $\alpha$-band. We find a good agreement between the experimental data and the calculated dispersion by introducing an interband SO coupling strength of $\Delta_{1,2}$ = 120 meV. 

In contrast to the surface alloy BiAg$_2$ investigated in Ref. \cite{Bentmann:2012} where the Rashba-type SOC is larger than the interband SOC ($\Delta \approx$ 30 meV, $E_{RB}= \alpha_{RB}^2m^{\star}/(2\hbar^2) \approx 300$~meV), the relative energy scale of SOC induced by the Rashba effect and by the interband coupling is inverted for Pb/Bi/Si. 
We find the interband contribution to be three orders of magnitude stronger than the Rashba contribution, $E_{RB} = 0.23$ meV. 
The larger $\Delta$ in Pb/Bi/Si(111) by a factor of four compared to the one found in BiAg$_2$ is understood by the fact that in Pb/Bi/Si a Rashba-split electron-like band interacts with a Rashba-split hole-like band. In BiAg$_2$ the interband SO coupling mixes electronic states which all disperse with m$^{\star} < 0$. The sign change of the momentum splitting for the $\alpha$-band is a direct spectroscopic evidence of the mixed and k-dependent (via $h_k$ and $g_k$) spin character of the corresponding new wave functions. 

In the following we corroborate that in Pb/Bi/Si the interband SO coupling occurs between bands with \emph{anti-parallel} spins (e.g. spin-flip term, $\Delta_{1,2}=\Delta^{\uparrow, \downarrow}$), instead of a coupling of bands with parallel spins (e.g. spin-conserving term, $\Delta^{\uparrow (\downarrow), \uparrow (\downarrow)}$). 
Although the latter scenario can also explain the modification of the $\alpha$-band dispersion (see dashed lines in Fig. \ref{Fig5}(b)) it fails to qualitatively reproduce the measured spin splittings as a function of momentum in the hybridization region, i.e. in the momentum region starting at the inflection point. A comparison between both models and the data is shown in Fig. \ref{Fig5}(c). The model of interband SO coupling of states with parallel spins predicts neither a decrease of the spin splitting, nor a vanishing spin splitting along the band dispersion. In contrast, the model of the coupling of states with \emph{anti-parallel} spins matches qualitatively the experimental data.
That the energy splitting between the $H^{+}$ and the $G^{+}$ branch does not decrease linearly with momentum may serve as an indication that the degree of the spin polarization of the new eigenstates ($|H^{+}\rangle$, $|G^{+}\rangle$), which are formed by mixing of the Rashba spinors with opposite spin directions, is no longer 100\%, although this has been assumed in the fit. Indeed, theory predicts that toward the avoided crossing point the spin polarization (P) of the states $|H^{\pm}\rangle$ and  $|G^{\pm}\rangle$ reduces according to \cite{Fabian:1998,Gradhand:2009}
\begin{eqnarray}
P(k_{\parallel})^{H^{\pm}} = \frac{(\alpha^{+}_{k_{\parallel}}-\beta^{-}_{k_{\parallel}})}{\sqrt{(\alpha^{+}_{k_{\parallel}}-\beta^{-}_{k_{\parallel}})^2+4\Delta^2}}, \\ P(k_{\parallel})^{G^{\pm}} = \frac{(\alpha^{-}_{k_{\parallel}}-\beta^{+}_{k_{\parallel}})}{\sqrt{(\alpha^{-}_{k_{\parallel}}-\beta^{+}_{k_{\parallel}})^2+4\Delta^2}}
\end{eqnarray}
Consequently the measured spin polarization of the $\alpha$-band toward the crossing point is reduced (i) due to the crossing of branches $H^{+}$ and $G^{+}$ and (ii) because the spin polarization of each branch is reduced due to the mixed spin character. Another indication of the action of the spin-flip term is that both bands $H^{+}$ and $G^{+}$ change their spin direction when passing the crossing point. If the spin-conserving term would dominate the spin character would not change when going along these bands \cite{Ebert:1997}. Whether the action of the spin-flip, or the spin-conserving term of the SOC operator is present in a system, depends on the details of the orbital symmetry of the interacting wave functions. For example, if the $\beta$-band were $p_z$ derived then $\Delta^{\uparrow, \downarrow} = 0$ and $\Delta^{\uparrow (\downarrow), \uparrow (\downarrow)} \neq 0$, i.e. the spin-conserving part would dominate.

\section{Pb/Cu(111)}
\begin{figure}[htb]
\begin{center}
\includegraphics[width=0.5\textwidth]{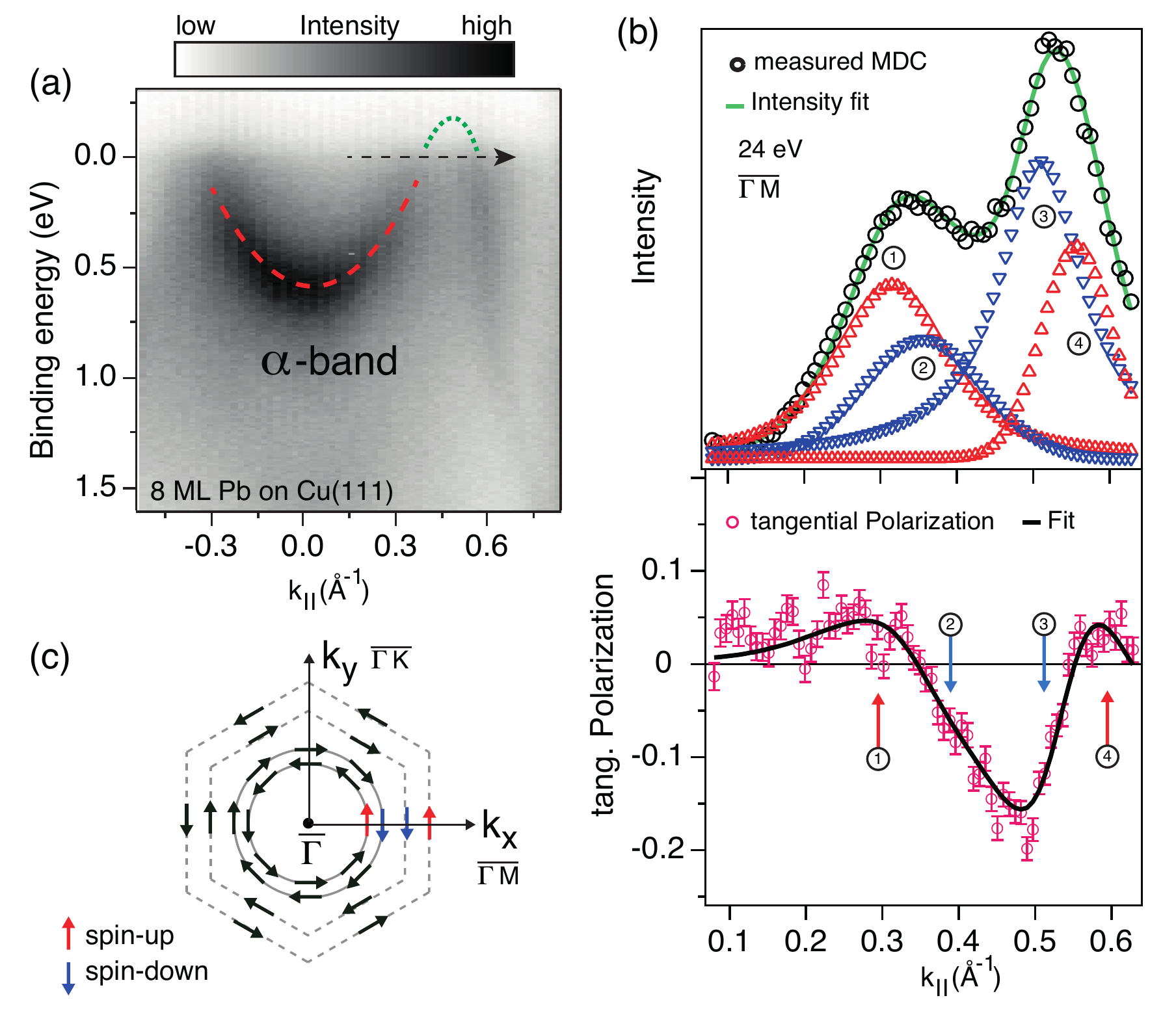}
\caption{(a) (color online) Band structure of a 8 ML thick Pb film on Cu(111). The dashed line is a fit of the k-dependent intensity maxima to quantify the effective mass. The dashed arrow indicates the location of the spin-resolved measurement. (b) (upper panel) Spin-integrated momentum distribution curve and decomposition into four peaks as revealed from the two-step fitting routine and (lower panel) tangential spin polarization data and fit. (c) Spin texture of the $\alpha$-band in Pb/Cu(111) at $E_F$.} 
\label{Fig7}
\end{center}
\end{figure}
In the following we substantiate that the interband SO coupling, which explains (i) the modification of the dispersion of the $\alpha$-band and (ii) the sign reversal in the momentum splitting, is independent of the supporting substrate by studying QWS in Pb/Cu(111) \cite{Dil:2004,Mathias:2010,Braun:2009,Zugarramurdi:2009,Zugarramurdi:2012,Otero:2002}.
Figure \ref{Fig7}(a) shows the electronic structure of 8 ML Pb on Cu(111). In contrast to the electronic structure of Pb/Bi/Si the effective mass of the $\alpha$-band is significantly reduced from $m^{\star}$ = 3.2 $m_e$ to 0.86 $m_e$, and the pronounced change in the dispersion is situated above the Fermi level. The enhanced dispersion allows to perform a SR-MDC indicated by the dashed arrow in Fig. \ref{Fig7}(a). The results and our analysis are displayed in Fig. \ref{Fig7}(b) where we show the measured spin-integrated MDC close to $E_F$ (upper panel) and the tangential spin polarization data (lower panel), respectively. We find a sign change in the momentum splitting of the $\alpha$-band which is consistent with previous results obtained from Pb QWS on the Bi reconstructed Si. This is further illustrated by comparing Fig. \ref{Fig7}(c), which displays the spin alignment at $E_F$, with the inset of Fig. \ref{Fig8}(b).
\section{Conclusions}
To summarize, we have identified a momentum region in Pb QWS where the description of the Rashba effect breaks down due to the interband spin-orbit coupling. Essential is that this coupling leads to a hybridization of Rashba states with \emph{anti-parallel} spins thereby causing the pronounced change in the dispersion of the $\alpha$-band, a sign change of the momentum splitting, and in particular a vanishing spin splitting toward the avoided crossing point. These modifications clearly reflect the SOC-induced hybridization mechanism because of the mixed spin and orbital character of the new wave functions.
This finding is consistent with the results obtained from QWS in Pb on Cu(111). Furthermore, we have found that the interband contribution in Pb/Bi/Si is three orders of magnitude stronger then the SOI contribution by the Rashba effect. It is expected that this ratio is smaller in  Pb/Pb/Si and larger in Pb/Ag/Si due to the different effective masses of the $\alpha$-band.

\vspace{0.5cm}
We thank F. Dubi, M. Kropf, and C. Hess for support during the measurements. This work was supported by the Swiss National Science Foundation.

\end{document}